\documentclass{aastex631}

\shorttitle{Approximating BH Spins}
\shortauthors{Bavera, Zevin, \& Fragos 2021}
\usepackage{amsmath}
\usepackage{graphicx}
\usepackage{txfonts}
\usepackage{xspace}
\def\donea{\ensuremath{0.059305}\xspace}
\def\dtwoa{\ensuremath{0.035552}\xspace}
\def\dthreea{\ensuremath{0.270245}\xspace}
\def\doneb{\ensuremath{0.026960}\xspace}
\def\dtwob{\ensuremath{0.011001}\xspace}
\def\dthreeb{\ensuremath{0.420739}\xspace}

\def\conea{\ensuremath{0.051237}\xspace}
\def\ctwoa{\ensuremath{0.029928}\xspace}
\def\cthreea{\ensuremath{0.282998}\xspace}
\def\coneb{\ensuremath{0.027090}\xspace}
\def\ctwob{\ensuremath{0.010905}\xspace}
\def\cthreeb{\ensuremath{0.422213}\xspace}

\definecolor{chmagenta}{rgb}{0.54, 0.17, 0.88}

\graphicspath{{./}{figures/}}
\begin{document}

\title{Approximations to the spin of close Black-hole--Wolf-Rayet binaries}

\newcommand{\KICP}{\affiliation{Kavli Institute for Cosmological Physics, The University of Chicago, 5640 South Ellis Avenue, Chicago, Illinois 60637, USA}}
\newcommand{\EFI}{\affiliation{Enrico Fermi Institute, The University of Chicago, 933 East 56th Street, Chicago, Illinois 60637, USA}}
\newcommand{\Geneva}{\affiliation{Departement d’Astronomie, Université de Genève, Chemin Pegasi 51, CH-1290 Versoix, Switzerland}}

\author[0000-0002-3439-0321]{Simone S. Bavera}\thanks{simone.bavera@unige.ch}
\Geneva

\author[0000-0002-0147-0835]{Michael Zevin}\thanks{NASA Hubble Fellow}
\KICP \EFI

\author[0000-0003-1474-1523]{Tassos\,Fragos}
\Geneva

%% Note that the \and command from previous versions of AASTeX is now
%% depreciated in this version as it is no longer necessary. AASTeX 
%% automatically takes care of all commas and "and"s between authors names.

%% AASTeX 6.31 has the new \collaboration and \nocollaboration commands to
%% provide the collaboration status of a group of authors. These commands 
%% can be used either before or after the list of corresponding authors. The
%% argument for \collaboration is the collaboration identifier. Authors are
%% encouraged to surround collaboration identifiers with ()s. The 
%% \nocollaboration command takes no argument and exists to indicate that
%% the nearby authors are not part of surrounding collaborations.

%% Mark off the abstract in the ``abstract'' environment. 
\begin{abstract}
Population synthesis studies of binary black-hole mergers often lack robust black-hole spin estimates as they cannot accurately follow tidal spin-up during the late black-hole--Wolf-Rayet evolutionary phase. 
We provide an analytical approximation of the dimensionless second-born black-hole spin given the binary orbital period and Wolf-Rayet stellar mass at helium depletion or carbon depletion. 
These approximations are obtained from fitting a sample of around $10^5$ detailed MESA simulations that follow the evolution and spin up of close black-hole--Wolf-Rayet systems with metallicities in the range $[10^{-4},1.5Z_\odot]$. 
Following the potential spin up of the Wolf-Rayet progenitor, the second-born black-hole spin is calculated using up-to-date core collapse prescriptions that account for any potential disk formation in the collapsing Wolf-Rayet star. 
The fits for second-born black hole spin provided in this work can be readily applied to any astrophysical modeling that relies on rapid population synthesis, and will be useful for the interpretation of gravitational-wave sources using such models. 
\end{abstract}

%% Keywords should appear after the \end{abstract} command. 
%% The AAS Journals now uses Unified Astronomy Thesaurus concepts:
%% https://astrothesaurus.org
%% You will be asked to selected these concepts during the submission process
%% but this old "keyword" functionality is maintained in case authors want
%% to include these concepts in their preprints.

%% From the front matter, we move on to the body of the paper.
%% Sections are demarcated by \section and \subsection, respectively.
%% Observe the use of the LaTeX \label
%% command after the \subsection to give a symbolic KEY to the
%% subsection for cross-referencing in a \ref command.
%% You can use LaTeX's \ref and \label commands to keep track of
%% cross-references to sections, equations, tables, and figures.
%% That way, if you change the order of any elements, LaTeX will
%% automatically renumber them.
%%
%% We recommend that authors also use the natbib \citep
%% and \citet commands to identify citations.  The citations are
%% tied to the reference list via symbolic KEYs. The KEY corresponds
%% to the KEY in the \bibitem in the reference list below. 

\section*{}

Isolated binary evolution is one of the leading astrophysical mechanisms proposed for generating merging binary black holes (BBHs).
In the standard BBH formation scenarios through isolated binary evolution, BBH progenitors either proceed through a stable mass transfer and a common envelope phase~\citep[e.g.,][]{Bethe1998,Belczynski2002,Kalogera2007,Dominik2012,Bavera2020} or a double stable mass-transfer episode~\citep[e.g.,][]{VandenHeuvel2017,Neijssel2019,Bavera2021}. 
In both cases, after the second mass-transfer event the binary emerges as a black-hole--Wolf-Rayet (BH-WR) system. 
The first-born BH spin is determined by the angular momentum (AM) transport of the progenitor star during the red supergiant evolutionary phase. 
Asteroseismic observations hint at efficient AM transport~\citep{Spruit1999,Spruit2002,Fuller2019a}, and hence, upon expansion, any initial AM is mostly transported to the outer layers of the star which are subsequently lost due to mass transfer and wind mass loss. 
The dimensionless spin parameter of the first-born BH is therefore $a_\mathrm{BH1}\simeq 0$~\citep{Fragos2015,Qin2018,Fuller2019b}. 
During the BH-WR evolutionary phase, if the binary orbit is tight enough, the WR star experiences tidal spin up from the compact-object companion, which can lead to rapid rotation and, after core collapse, to the formation of a rapidly rotating BH~\citep{Qin2018,Bavera2020,Bavera2021}. 
However, during the BH-WR evolution, mass loss through metallicity-dependent stellar winds can widen the binary and might lead to tidal decoupling. 
A careful treatment of binary interactions, as well as the WR stellar structure evolution and core collapse, are thus essential to properly determine the second-born BH spin $a_\mathrm{BH2}$. 
In this research note, we present an analytical model to approximate $a_\mathrm{BH2}$ given the BH-WR orbital period and WR mass at helium (He) or carbon (C) depletion. 
Though the BH-WR orbital period and WR mass are implicitly dependent on BH companion mass and the zero age He main sequence (ZAHeMS) metallicity, we find that accurate fits for $a_\mathrm{BH2}$ can be obtained using this two-parameter model. 
This approximation is well suited for isolated binary evolution population synthesis studies aiming to study merging BBHs as well as modeling of other formation channels that rely on rapid population synthesis. 

To approximate $a_\mathrm{BH2}$, we use the results of around $10^5$ detailed BH-WR \texttt{MESA} simulations presented in \cite{Bavera2021}. 
These simulations take into account differential stellar rotation, tidal interaction, WR stellar winds, the evolution of WR stellar structure, and up to date core-collapse prescriptions accounting for AM supported disk formation during collapse which guarantees the general relativistic limit $a_\mathrm{BH2} \leq 1$. 
The $a_\mathrm{BH2}$ approximation presented in this work is therefore model dependent on the assumptions made, which include \cite{Fryer2012} delayed core-collapse mechanism, fit to the (pulsational) pair-instability supernovae models of \cite{Marchant2019}, up to $0.5\,\mathrm{M}_\odot$ neutrino mass and AM loss as implemented in \cite{Zevin2020b}, and the arbitrary minimum BH mass of $2.5\,\mathrm{M}_\odot$.
The initial conditions of the simulations at ZAHeMS cover the range of BH-WR orbital periods $p^i \in [0.09,8]\,\mathrm{day}$, WR masses $m_\mathrm{WR}^i \in [8,80]\,\mathrm{M}_\odot$, BH masses $m_\mathrm{BH1}^i\in[2.5,55]\,\mathrm{M}_\odot$, and metallicities $Z^i\in[10^{-4},1.5Z_\odot]$, all of which are uniformly sampled in log. 
All assumptions about these simulations are explain in detail in Appendixes C and D of \cite{Bavera2021}. The simulation dataset used in this work is made available, including the second-born BH mass $m_\mathrm{BH2}$.

\begin{figure}
\centering
\includegraphics[width=\linewidth]{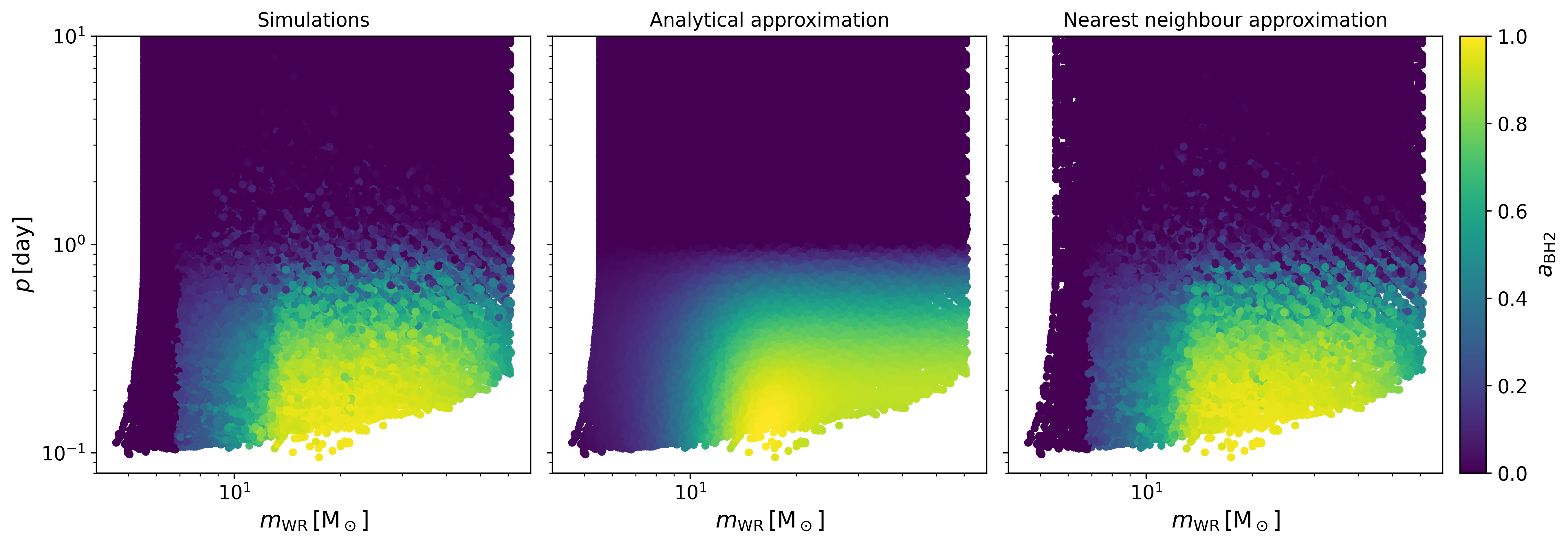} 
\caption{
Second-born black hole spin of close BH-Wolf-Rayet systems as a function of binary orbital period and Wolf-Rayet mass at carbon depletion. 
\textit{Left:} The result of the $\sim10^5$ post-processed \texttt{MESA} simulation for all BH companion masses ($m_\mathrm{BH1}\in[2.5,55]\,\mathrm{M}_\odot$) and metallicities ($Z\in[10^{-4},1.5Z_\odot]$). 
\textit{Center:} Analytical approximation for the second-born BH spin presented in this work (Eq. \ref{eq:abh}-\ref{eq:f}). 
\textit{Right:} Nearest neighbour approximation for the second-born BH spin determined via performing 20 realizations where 5\% of the original dataset is used to test the nearest neighbor algorithm. 
}
\label{fig:R_fiducial}
\end{figure}

We empirically determine that $a_\mathrm{BH2}$ can be approximated by the following analytical function, given the BH-WR orbital period, $p$, and Wolf-Rayet mass, $m_\mathrm{WR}$, at He or C depletion: 
\begin{equation}
    a_\mathrm{BH2} =
    \begin{cases}
        \alpha \log_{10}(p/[\mathrm{day}])^2+\beta \log_{10}(p/[\mathrm{day}]) & p \leq 1 \,\mathrm{day} \\
        0 & \text{otherwise}
    \end{cases}   
\label{eq:abh}
\end{equation}
where $\alpha=f(m_\mathrm{WR}, c_1^\alpha, c_2^\alpha, c_3^\alpha)$ and $\beta=f(m_\mathrm{WR}, c_1^\beta, c_2^\beta, c_3^\beta)$ for
\begin{equation}
    f(m_\mathrm{WR}, c_1, c_2, c_3) = \frac{-c_1}{c_2+\exp(-c_3 \, m_\mathrm{WR}/[\mathrm{M}_\odot])} \, .
\label{eq:f}
\end{equation}
The coefficients $c^{\alpha,\beta}_{1,2,3}$ are determined through non-linear least-square minimization to be $(c_1^\alpha,c_2^\alpha,c_3^\alpha)=(\donea,\dtwoa,\dthreea)$ and $(c_1^\beta,c_2^\beta,c_3^\beta) = (\doneb,\dtwob,\dthreeb)$ for input $p$ and $m_\mathrm{WR}$ at He depletion, or $(c_1^\alpha,c_2^\alpha,c_3^\alpha)=(\conea,\ctwoa,\cthreea)$ and $(c_1^\beta,c_2^\beta,c_3^\beta)=(\coneb,\ctwob,\cthreeb)$ for input $p$ and $m_\mathrm{WR}$ at C depletion. 
Though our simulations and analytical approximation always return $0 \leq a_\mathrm{BH2} \leq 1$, one may wish to impose the extra condition $a'_\mathrm{BH2}=\max(a_\mathrm{BH2},1)$ to ensure that BHs are spinning at the physical limit according to general relativity if the fit is extrapolated to regimes outside the parameter space of our models. 

We compare the absolute errors $|a^{true}_\mathrm{BH2}-a^{approx.}_\mathrm{BH2}|$ from our analytical approximation to the absolute errors determined by $k$-fold validation of a nearest neighbor interpolant trained on the normalized input parameters $(\log_{10}(p),\log_{10}(m_\mathrm{WR}))$. 
We find the absolute errors to be comparable: over the whole population for all orbital periods (orbital periods less than 1 day), the absolute error of the analytical approximation is $0.00^{+0.13}_{-0.00}$ ($0.04^{+0.14}_{-0.04}$) whereas the absolute error of the nearest neighbor algorithm is $0.00^{+0.09}_{-0.00}$ ($0.02^{+0.12}_{-0.02}$), where we quote the median and symmetric 90\% credible interval. 
%find that this analytical approximation for $a_\mathrm{BH2}$ results in absolute errors $|a^\mathrm{true}_\mathrm{BH2}-a^{approx.}_\mathrm{BH2}|$ comparable to nearest neighbour appreciation as presented in Table~\ref{tab:AE}. 
%Here, the nearest neighbour interpolation is conducted after log-scaling the input parameter space $(p,m_\mathrm{WR})$ followed by a normalisation to the range $[0,1]$. 
A visual comparison between the simulation results, analytical approximation, and nearest neighbour algorithm is shown in Figure~\ref{fig:R_fiducial}. 
The accuracy of the two-parameter approximation to the full 4D parameter space $(p,m_\mathrm{WR},m_\mathrm{BH1},Z)$ is achieved due to the fact that the orbital period $p$ at He  or C depletion is itself a function of $m_\mathrm{WR}$, $m_\mathrm{BH1}$, and $Z$. 
Approximately, because for ZAHeMS $p^i < 1$ day the BH-WR evolutionary timescale of a few 100 kyr is larger than the tidal synchronisation timescale \citep{Qin2018}, to prove this point, one can assume that at He or C depletion any binary undergoing tidal spin-up is tidally synchronized. 
One therefore finds $a_\mathrm{BH2}=cJ_\mathrm{WR}/Gm_\mathrm{WR}^2 \propto \Omega / m_\mathrm{WR}^2$ for tidally spun up BHs, where $\Omega \propto 1/p$ is the orbital frequency. 
On the other hand, for wide BH-WR binaries, here $p\geq1$ day, it holds $cJ_\mathrm{WR} << Gm_\mathrm{WR}^2$ and hence $a_\mathrm{BH2}=0$. 

We anticipate that the flexible analytical approximation presented in this work will be useful for various population modeling endeavors. 
Spins are one of the key observational signatures of BBHs detected via gravitational waves, and an accurate physical representation of BH spins from theoretical models is essential for the interpretation of gravitational-wave events.

% \begin{table}
% \centering
% \begin{tabular}{ | c | c | c | }
% \hline
% & at He-depletion & at C-depletion \\
% \hline
% $c_1^a$ & \donea & \conea  \\ 
% $c_2^a$ & \dtwoa & \ctwoa \\  
% $c_3^a$ & \dthreea & \cthreea \\
% $c_1^b$ & \doneb & \coneb \\ 
% $c_2^b$ & \dtwob & \ctwob \\  
% $c_3^b$ & \dthreeb & \cthreeb \\
% \hline
% \end{tabular}
% \caption{Coefficients for the analytical approximation (Eq.~\ref{eq:f}) of the second-born BH at helium (He) and carbon (C) depletion.}
% \label{tab:coeff}
% \end{table}

% \begin{table}
% \centering
% \begin{tabular}{ | c | c | c | c | c |}
% \hline
% & analy. approx. (He/C-dep.)  & NN approx. (He/C.-dep.) & analy. approx. (He/C-dep.)  & NN approx. (He/C.-dep.)  \\
% \hline
%  & any $p$ & any $p$ & $p\leq 1$ day &  $p\leq 1$ day \\
% mean & 0.03 & 0.02 & 0.06 & 0.04  \\
% median & 0.00 & 0.00 & 0.04 & 0.02 \\
% 68\% CI & [0.00, 0.06] & [0.00,0.04] & [0.01, 0.11] & [0.00, 0.07] \\
% 90\% CI & [0.00, 0.13] & [0.00,0.09] & [0.00,0.18] & [0.00, 0.14]  \\
% \hline
% \end{tabular}
% \caption{Absolute errors, $|a^\mathrm{true}_\mathrm{BH2}-a^{approx.}_\mathrm{BH2}|$, of the analytical approximation versus the nearest neighbour approximation. For the nearest neighbour approximation we quote the average result from 20 tests against a sample size of 5\% the original one.}
% \label{tab:AE}
% \end{table}

\section{Data release statement}
We provide the dataset of BH-WR binaries at He and C depletion used in the work as well as a notebook with our analytical models on GitHub \url{https://github.com/ssbvr/approximating_BH_spins}. 
The \texttt{MESA} inlists used to run the BH-WR simulations are available at \url{http://cococubed.asu.edu/mesa_market/} or on Zenodo~\citep{SpinFit_Dataset}.

\vspace{0.5cm}
%\begin{acknowledgements}
\noindent We would like to thank Christopher Berry and Emmanouil Zapartas for comments on the manuscript. This work was supported by the Swiss National Science Foundation Professorship grant (project number PP00P2\_176868). Support for M.Z. was provided by NASA through the NASA Hubble Fellowship grant HST-HF2-51474.001-A awarded by the Space Telescope Science Institute, which is operated by the Association of Universities for Research in Astronomy, Inc., for NASA, under contract NAS5-26555. This study made use of the following open-sources Python modules \texttt{Matplotlib}~\citep{matplotlib}, \texttt{Numpy}~\citep{numpy3}, \texttt{Scikit-learn}~\citep{scikit-learn}, and \texttt{Scipy}~\citep{scipy}.
%\end{acknowledgements}

\bibliography{main}{}
\bibliographystyle{aasjournal}

%% This command is needed to show the entire author+affiliation list when
%% the collaboration and author truncation commands are used.  It has to
%% go at the end of the manuscript.
%\allauthors

%% Include this line if you are using the \added, \replaced, \deleted
%% commands to see a summary list of all changes at the end of the article.
%\listofchanges

\end{document}